\documentclass{aa}  

\makeatletter
\AtBeginDocument{%
	\geometry{
		hmargin=1.8cm
	}%
}

\let\AA@old@journalname\aa@journalname
\let\AA@old@manuscriptname\aa@manuscriptname
\let\AA@old@AALogo\AALogo
\let\AA@old@today\today
\newlength\AA@old@fboxrule
\newlength\AA@old@fboxsep

\newcommand*\AA@disableTitleBanner{%
	\renewcommand*\aa@journalname{}% remove "Astronomy & Astrophysics"
	\renewcommand*\aa@manuscriptname{}% remove "manuscript no. ..."
	\renewcommand*\AALogo{}% remove the logo
	\def\today{}% remove the date (used in the banner)
	\setlength{\AA@old@fboxrule}{\fboxrule}% hide the frame
	\setlength{\AA@old@fboxsep}{\fboxsep}%
	\setlength{\fboxrule}{0pt}%
	\setlength{\fboxsep}{0pt}%
}

\newcommand*\AA@restoreTitleBanner{%
	\let\aa@journalname\AA@old@journalname
	\let\aa@manuscriptname\AA@old@manuscriptname
	\let\AALogo\AA@old@AALogo
	\let\today\AA@old@today
	\setlength{\fboxrule}{\AA@old@fboxrule}%
	\setlength{\fboxsep}{\AA@old@fboxsep}%
}

\let\AA@old@maketitle\maketitle
\renewcommand*\maketitle{%
	\AA@disableTitleBanner
	\AA@old@maketitle
	\AA@restoreTitleBanner
}

\makeatother

\usepackage{fancyhdr}
\fancypagestyle{plain}{%
	\fancyhf{}                 % clear everything
	\fancyfoot[C]{\thepage}    % keep centered page number
	
}
\usepackage[utf8]{inputenc}
\usepackage[varg]{txfonts}  % A&A recommended font
\usepackage{bm}
\usepackage{amssymb}
\usepackage{amsmath}
\usepackage{mathrsfs}
\usepackage{graphicx}
\usepackage{tabularx}
\usepackage{enumitem}
\usepackage{color}
\usepackage{subcaption}
\usepackage{hyperref}
\usepackage{booktabs}
\usepackage{multirow}
\usepackage{array}

\usepackage{upgreek}
\usepackage{appendix}

\usepackage{natbib}

\hypersetup{
	colorlinks = true,
	linkcolor = blue,
	citecolor = blue,
	filecolor = blue,
	urlcolor = blue
}

\def\apjl{ApJL}
\def\apjs{Astrophys. J. Supp.}

\def\aap{Astron. Astrophys.}

\newcolumntype{K}[1]{>{\centering\arraybackslash}p{#1}}

\let\OLDthebibliography\thebibliography
\renewcommand\thebibliography[1]{
	\OLDthebibliography{#1}
	\setlength{\parskip}{0pt}
	\setlength{\itemsep}{0pt plus 0.3ex}
}

\allowdisplaybreaks

\titlerunning{Spin-mass correlation and VRR--GW231123}

\title{Black hole spin-mass correlation and vector resonant relaxation in gaseous star clusters: the origin of GW231123?}

\author{Zacharias Roupas\inst{1,2}}

\institute{Dipartimento di Fisica ``G. Occhialini'', 
	Universit\'a degli Studi di Milano-Bicocca, Piazza della Scienza 3, 20126 Milano, Italy
	\email{zacharias.roupas@unimib.it}
	\and
	Istituto Nazionale di Fisica Nucleare (INFN), Sezione di Milano-Bicocca, 
	Piazza della Scienza 3, 20126 Milano, Italy
}

\date{}

\abstract{
During the formation of a star cluster, a spin-mass correlation among stellar black holes is generated as they grow by accreting residual gas. Moreover, the black hole spin tends to be anti-aligned with its orbital angular momentum in the cluster. We show that GW231123, reported by the LIGO-Virgo-KAGRA (LVK) collaboration, lies on our predicted high-mass, high-spin plateau of the spin-mass correlation with positive Bayesian evidence over the LVK prior and the characteristic spin signature of the hierarchical-merger channel, coupled with very strong evidence favoring it over the GWTC-5.0 population-informed fiducial model. Furthermore, vector resonant relaxation (VRR) equilibrium is favored over the isotropic LVK prior in terms of reproducing the distribution of the relative spin tilt. The joint Bayes factor suggests strong evidence to support the favored cluster models. GW231123 is thus consistent with our proposed channel that generates correlated black hole masses and spins, while driving the spin orientations. 
}


\begin{document}
	
	\maketitle
	
\section{Introduction}\label{sec:intro}
	
The event GW231123 is the most massive binary black hole (BBH) merger observed by the LIGO-Virgo-KAGRA (LVK) collaboration,
with source-frame component masses of $137_{-17}^{+22} \,{\rm M}_\odot$ and $103_{-52}^{+20} \,{\rm M}_\odot$, and corresponding spins of $0.9_{-0.19}^{+0.10}$ and $0.8_{-0.51}^{+0.20}$, at a redshift of 
$0.40^{+0.27}_{-0.25}$ \citep{2025ApJ...993L..25A}. The primary black hole (BH) lies within or above the
theorized pair-instability mass gap, $\approx 60\text{--}130\,M_\odot$ \citep{2019ApJ...887...53F,2021ApJ...912L..31W,2023MNRAS.526.4130H,2026Natur.652..874T}, while the
secondary spans this range. These masses, together with the high spins, challenge attempts to identify the astrophysical origin of the signal.
Proposed scenarios include dynamical formation through hierarchical mergers \citep{2025ApJ...994L..54P,2026ApJ...999..127L,2026ApJ...999..236P}, 
population III stars \citep{2025ApJ...993L..54G,2025ApJ...993L..30L,2026ApJ..1003...80T},
gas accretion \citep{2025A&A...702A.208R,2026A&A...709A.5R,2026ApJ...996L..44B}, 
primordial BHs \citep{2025PhRvD.112h1306Y,2026PhRvL.136t1401D}, chemically homogeneous stellar evolution \citep{2025ApJ...995L..76P},
and direct core collapse \citep{2026MNRAS.546ag073C}.

We suggest that the BH masses, the BH spins, and the redshift of GW231123 could be explained by the BH mass-growth mechanism in gaseous star clusters we proposed in another recent work
\citep{2025A&A...702A.208R,2026A&A...709A.5R}. The additional hypothesis of vector resonant relaxation \citep{1996NewA....1..149R,2017ApJ...842...90R}, combined with the accretion-disk formation mechanism we also recently proposed \citep{2026arXiv_Roupas}, explains the relative BH spin tilt. Those theoretical elements define a BBH formation channel with properties resembling those of (but not restricted to) GW231123 and they are testable against similar signals in the future.

Specifically, in \cite{2026A&A...709A.5R} we identified a spin-mass correlation of BHs, generated by accretion in proto-clusters. Stellar BHs, originating from the most massive stars, are born so early in the life of a sufficiently compact star cluster that they have enough time to accrete the residual gas, following the first star formation event, before the gas gets depleted by stellar winds and supernova (SN) explosions
\citep{2019A&A...632L...8R,2025A&A...702A.208R}. Their mass can grow via accretion to values within the mass gap and up to $\approx 10^3\,{\rm M}_{\odot}$ \citep{2025A&A...702A.208R}, while they also gain spin \citep{2026A&A...709A.5R}. As the Bondi sphere is advected along the BH orbit, angular momentum is injected into the sphere, induced by the velocity shear between the inner and outer hemispheres, driving a BH spin anti-alignment with the orbital angular momentum
\citep{2026arXiv_Roupas}. Therefore, the distribution of orbital orientations of the BHs is reflected in the relative spin orientations of a BBH formed from those BHs. Such a distribution can be provided by the process of vector resonant relaxation (VRR), which can operate in the cores of sufficiently compact clusters either if an intermediate-mass black hole (IMBH) is formed and driven in the cluster center or even in the absence of a central IMBH
\citep{2019ApJ...878..138M,2020JPhA...53d5002R}.

In the next section, we describe how we computed the Bayesian evidence of the spin-mass correlation hypothesis for GW231123. In Section \ref{sec:vrr}, we show that the VRR hypothesis has the capacity to generate the observed relative spin tilt of GW231123 and calculate the joint Bayesian factor. In Section \ref{sec:time} we verify that the formation and merger timescales are consistent with the observed redshift.	We present and discuss the results of the analysis in Section \ref{sec:results} and present our conclusions in the final section \ref{sec:conclusion}.	 
	
\section{Spin-mass correlation}\label{sec:spin-mass}

\begin{figure}[tb]
	\centering
	\includegraphics[width=\columnwidth]{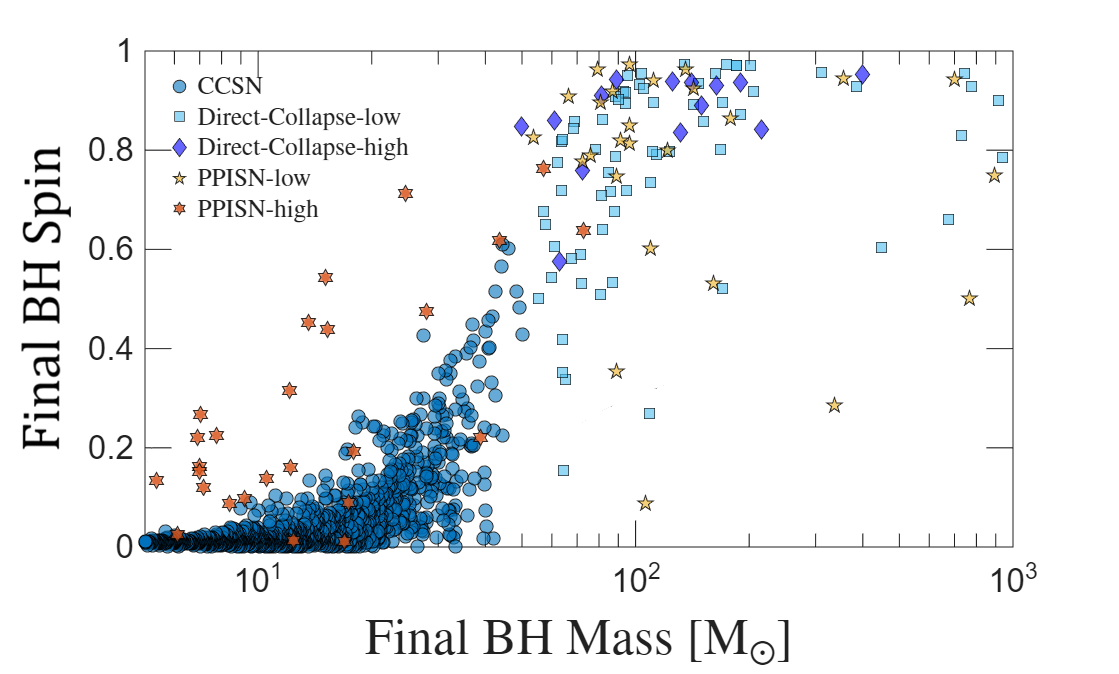}
	\caption{Final BH spin with respect to final BH mass of all BHs for one representative simulation run of our typical cluster with $M_{\star} = 10^6\,{\rm M}_\odot$, $r_{c,\star} = 1 \,{\rm pc}$, $Z=0.01\,{\rm Z}_{\odot}$. The legend labels provide the BH origin: CCSN for a core-collapse-supernova remnant; direct-collapse-low (high) for a BH generated by direct collapse of the core of a low- (high-) mass star; PPISN-low (high) for a BH remnant from a pulsational-pair-instability supernova of low- (high-) mass. From the simulations of \citet{2026A&A...709A.5R}.}
	\label{fig:a_m_fin_data}
\end{figure}

In the accretion-driven channel investigated in this work \citep{2026A&A...709A.5R}, the BH growth and spin-mass correlation are an integral part of the formation process of sufficiently massive and dense star clusters, especially at subsolar $\sim 0.1\, {\rm Z}_{\odot}$ and low $\lesssim 0.01\, {\rm Z}_{\odot}$ metallicities. Such proto-globular clusters have been observed at high redshift $\sim 10$ with the James Webb Space Telescope in the Cosmic Gems arc galaxy with typical masses $\sim 10^6\,{\rm M}_{\odot}$ and sizes $\sim 1\,{\rm pc}$ \citep{2024Natur.632..513A}. The semi-analytic model and simulation setup which follow the BH spin-up and mass growth are described in detail in
\citet{2025A&A...702A.208R, 2026Parti...9...18R} and \citet{2026A&A...709A.5R}. We briefly summarize it below. 

The gas and stellar components of the star cluster follow time-dependent Plummer profiles, while the model accounts for the cluster expansion driven by gas mass loss, the larger spatial extent of the gas relative to the stars, the gas temperature profile, and the gas cooling. 
The gas becomes depleted exponentially with a timescale calculated explicitly by estimating the stellar winds and supernova energy injection. Each run terminates once $99\%$ of the initial gas mass has been expelled. Each BH is created at the time set by the lifetime of its zero-age main sequence progenitor and is assigned the related generation process (direct collapse, core collapse supernova, pair-instability supernova) through the ZAMS-to-BH mass correspondence of \citet{2017MNRAS.470.4739S}. The initial BH masses are  drawn from a Salpeter mass function with an upper cut-off at $55\,{\rm M}_{\odot}$. Their initial positions and velocities are sampled from the gravitational Brownian probability density \citep{2021A&A...646A..20R}. 
The BH orbits are integrated with a drift-kick scheme. The deterministic forces of gravity and dynamical friction are evaluated with a Runge-Kutta method, using the Chandrasekhar formula for the stellar component, its hydrodynamic counterpart for the gas, and the deceleration due to the mass increase. Stochastic velocity increments are applied at adaptive timesteps, in practice $\Delta t \simeq 10^{-4}-10^{-2}\,{\rm Myr}$, modeled as white noise for the stellar encounters and as an Ornstein--Uhlenbeck process for the gas turbulence. 
The accretion is of isotropic Bondi--Hoyle type, with the accretion radius taken as the smaller of the Bondi--Hoyle and tidal-lobe radii at each instant, and the accretion rate capped at ten times the reference Eddington rate. The BH accretion disks are disrupted at each BH stochastic kick and the model accounts for both thin and slim disks depending on the accretion rate.

We elaborate further on the physics and major physical parameters in the following. The BHs, being the remnants of massive stars, are generated early in the life of the cluster, $\lesssim 2\,{\rm Myr}$, when residual gas from the first star formation event is still present. Lower gas metallicities imply weaker stellar winds which may provide a sufficient time-window for the early-formed BHs to accrete gas before it gets depleted by winds and supernova explosions.
For the sufficiently massive and dense clusters that can generate BHs with masses $\sim 10^2-10^3\,{\rm M}_{\odot}$, the gas-depletion timescale is still relatively short, with $99\%$ of the gas being depleted after $\sim 13\,{\rm Myr}$ \citep{2025A&A...702A.208R}. 
For this initial time of a cluster life, our simulations follow the dynamical evolution of BHs with initial masses $\leq 55\,{\rm M}_{\odot}$ subject to random fluctuations of the gravitational field which cause both reorientation of the BH spin relative to the accretion disk and the dislocation of the BH from its current trajectory. The lighter BHs are more sensitive to those perturbations, while the more massive BHs can form more persistent disks and accumulate closer to the center of the cluster where the gas density is higher. As a result, a saturating spin--mass relation is generated with low spins at low masses, rising continuously with mass through all intermediate spin values, to high spins at high masses, depicted in Fig.~\ref{fig:a_m_fin_data} for a typical simulation run. 
Throughout this work, $\chi \equiv c S/(G m^2) \in [0,1]$ denotes the dimensionless spin magnitude of a BH of mass $m$ and spin angular momentum, $S$.
In the simulations used here from \citet{2026A&A...709A.5R} we adopted initial BH spins of $\chi_{\rm ini}=0.01$, consistent with the recent models of efficient angular momentum transport during core contraction of the massive star progenitor (revised Tayler instability) \citep{2019ApJ...881L...1F,Fuller_2019,2020A&A...636A.104B}. 
Even in the case of natal spin $0.1$, consistent with the classic Tayler-Spruit dynamo \citep{Spruit_2002,Heger_2005,Qin_2018}, the general form of Fig.~\ref{fig:a_m_fin_data} persists with $\chi$ only shifted to values $\sim 0.1$ or lower for $m\lesssim 20\,{\rm M}_{\odot}$. The maximum spin, and the transition region remain nearly identical to our adopted $\chi_{\rm ini} = 0.01$ case. We considered cluster masses $M_{\star} = 7\cdot 10^5,\, 10^6\,{\rm M}_{\odot}$, sizes $r_c=0.7-1.2\,{\rm pc}$, star formation efficiency $0.35$, and metallicities $Z=0.1,\,0.01\,Z_{\odot}$, consistent with the Cosmic Gems clusters, and which can generate at least one BH with mass $\sim 150\,{\rm M}_{\odot}$ spanning the upper limit of the GW231123 primary component. 
Our simulations give in effect thin to slim disks with dimensionless accretion rate (BH accretion rate over Eddington rate, see \citealt{2026A&A...709A.5R}) $\dot{m} \sim 0.1-10$ depending on the gas density, along with the location and velocity of the BH. A BH with initial mass $50\,{\rm M}_{\odot}$ and $\dot{m}=5$ reaches $150\,{\rm M}_{\odot}$ within $\sim 6\,{\rm Myr}$. 
The spin (Eq.~(18) of \citealt{2026A&A...709A.5R}) reaches $\chi\simeq 0.9$ once the mass has grown by a factor $1.6$ in the thin-disk limit \citep{1970Natur.226...64B}, and $\simeq 1.8$ for the slim disk at $\dot{m}=5$, that is at $\sim 3\,{\rm Myr}$, well before the final mass is assembled. The timescales and the spin-mass relation follow from the accretion and gravitational physics implemented in the simulations, and they are not calibrated to any prescribed observable.

\begin{figure}[tbp]
	\centering
	\includegraphics[width=0.92\columnwidth]{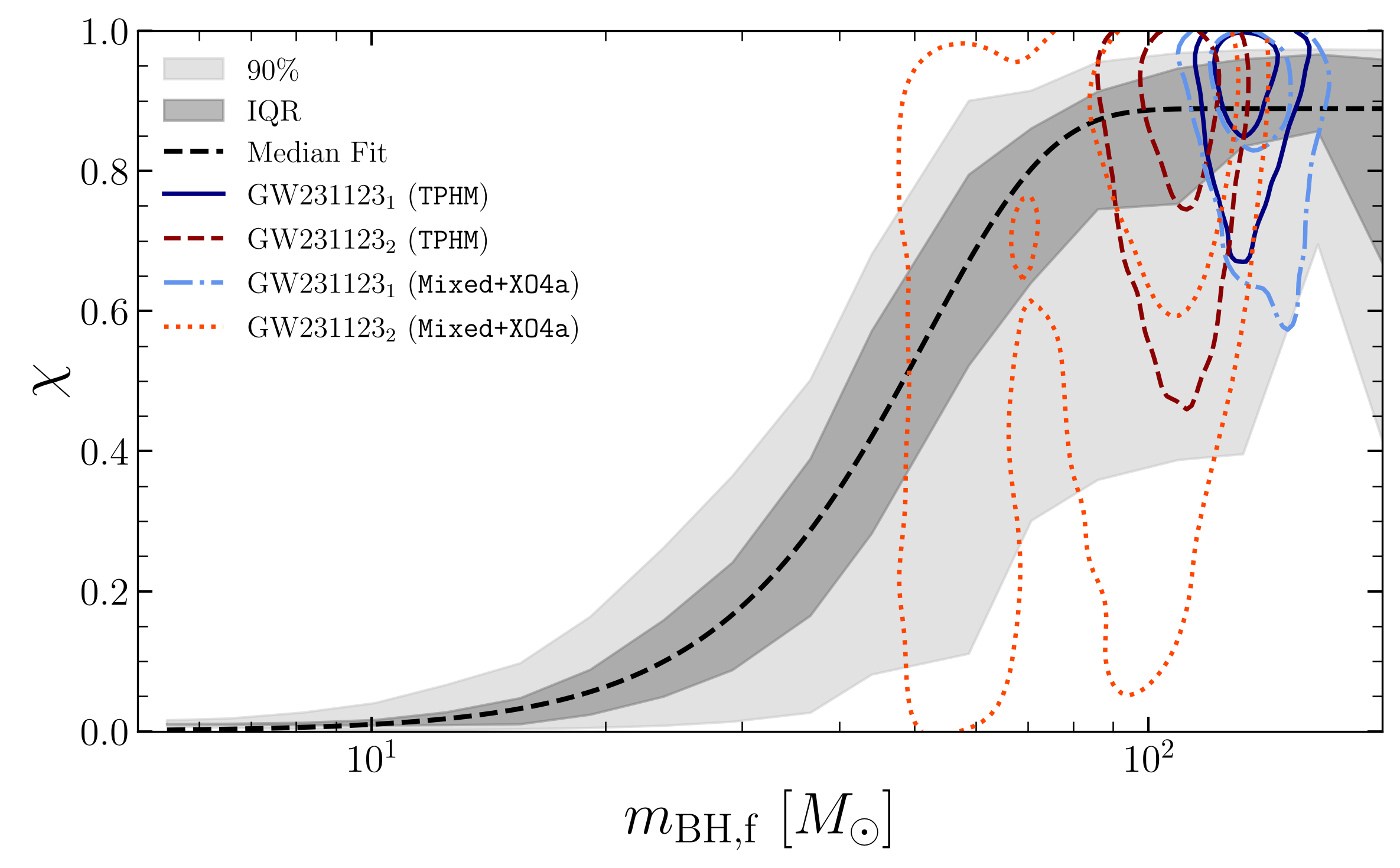}
	\caption{Accretion-driven spin-mass distribution of BHs predicted by \citet{2026A&A...709A.5R} $-$ median fit, interquartile range (IQR), and $90\%$ band $-$ for our typical cluster ($M_{\star} = 10^6\,{\rm M}_{\odot}$, $r_c=1\,{\rm pc}$, $Z=0.01\,{\rm Z}_{\odot}$), overlaid with the $50\%$ and $90\%$ credible contours of the two GW231123 components for the indicative \texttt{TPHM} waveform model and the combination of all waveform models, \texttt{Mixed+XO4a}. The broad, bimodal $90\%$ contour for the secondary component of the latter model reflects the much lower secondary mass and the uninformative secondary spin of \texttt{XO4a}.}
	\label{fig:spinmass}
\end{figure}

On the GW231123 side, we remark that the LVK analysis reports substantial waveform-model systematics \citep{2025ApJ...993L..25A}, with model-dependent differences in several inferred parameters.  
The waveform models \texttt{IMRPhenomXPHM} (\texttt{XPHM}) and \texttt{IMRPhenomXO4a} (\texttt{XO4a}) add nearly no information on the secondary spin, with a wide $90\%$ credible interval $\Delta_{90} \chi_2 = 0.86$, $0.88$, respectively, and small Hellinger distances \citep{Hellinger1909,LeCam1986} between posterior and prior distributions, $d_{{\rm H},i}^2 \equiv  1 - \int \sqrt{ p(\chi_i \mid d_{\rm GW})\, \pi_{\rm PE}(\chi_i) }\, {\rm d}\chi_i$. We obtained  $d_{{\rm H},2} = 0.17$, $0.07$, respectively, where $d_{\rm H} = 0$ for coinciding distributions, $d_{\rm H} = 1$ for disjoint supports, and $d_{\rm H} \simeq 0.34$ for two Gaussians offset by one standard deviation. In a spin--mass plot the secondary spin contour of those two models would span the whole range $0\leq \chi_2 \leq 1$ following the prior, rather than any measurement. We note that the median secondary mass of \texttt{XO4a} is $m_2\simeq 55\,{\rm M}_{\odot}$, nearly half the value inferred by the other models. The precessing-sector
phenomenology of \texttt{XO4a} is moreover calibrated to single-spin precessing simulations with $\chi_1 <0.8$, below the values inferred here. The remaining waveform models, \texttt{IMRPhenomTPHM} (\texttt{TPHM}), \texttt{SEOBNRv5PHM} (\texttt{V5PHM}), and \texttt{NRSur7dq4} (\texttt{NRSUR}), are significantly more informative with $\Delta_{90}\chi_2 \leq 0.64$ and $d_{{\rm H},2} \geq 0.36$ throughout; therefore we report the results for each one of these three models individually and for their equal-weight pool. We also report the equal-weight combination of all five waveform models released by the LVK as \texttt{Mixed+XO4a}, which was used to announce GW231123 \citep{2025ApJ...993L..25A}.
 
 Figure~\ref{fig:spinmass} shows that the GW231123 posterior contours are in striking agreement with the high-mass, high-spin plateau predicted by the \citet{2026A&A...709A.5R} model of BHs grown via gas accretion in proto-clusters.  In particular, the interquartile (IQR) of both primary and secondary components overlap the region in which the model predicts large spin magnitudes, $\chi \sim 0.9$, for BH masses on the order of $10^2\,M_\odot$. 
 
We quantify this apparent agreement with the conditional spin--mass Bayes factor of the accretion-driven spin-mass correlation hypothesis against a comparative model, $X$,
\begin{equation}\label{eq:bayes}
	\mathcal{B}^{X}_{\chi|m} =
	\frac{\left\langle p_{\rm corr}(\bm{\chi}\mid\bm{m})/
		\pi_{\rm PE}(\bm{\chi})\right\rangle}
	{\left\langle p_{X}(\bm{\chi}\mid\bm{m})/
		\pi_{\rm PE}(\bm{\chi})\right\rangle},
\end{equation}
where $\bm{\chi}=(\chi_1,\chi_2)$, $\bm{m}=(m_1,m_2)$, $\pi_{\rm PE}$ is the parameter-estimation prior of the event and both averages are taken over the same posterior samples released by the LVK Collaboration for GW231123 \citep{2025ApJ...993L..25A}. Since no joint spin prior has been released, we take $\pi_{\rm PE}(\bm{\chi}) = \pi_{\rm PE}(\chi_1)\pi_{\rm PE}(\chi_2)$, built from the released prior samples of each component, which are uniform over the allowed spin interval.

Under the accretion hypothesis, the spin magnitudes follow the mass-dependent density, 
$p_{\rm corr}(\bm{\chi}\mid\bm{m}) = p_{\rm corr}(\chi_1\mid m_1)
p_{\rm corr}(\chi_2\mid m_2)$, 
inferred from the \citet{2026A&A...709A.5R} simulations, where $m$ denotes the source-frame component mass. The simulated BHs are partitioned into logarithmically spaced final-mass bins. We estimated $p_{\rm corr}$ on the same bins in two ways. In the \textsc{Beta} estimator, we fit within each bin $b$ a Beta
distribution, ${\rm Beta}(\alpha_b,\beta_b)$, to the spin values by maximum likelihood, whose shape parameters determine the mean, width and skewness of the distribution, setting 
$p_{\rm corr}(\chi\mid m) = {\rm Beta}(\chi;\alpha_b,\beta_b)$ 
for a posterior sample with mass $m$ falling in bin $b$. The Beta distribution is a naturally bounded parametric choice for a spin magnitude, $\chi\in[0,1]$ and it allows for skewness and variable widths across mass bins. In the \textsc{Bin} estimator, we can calculate the density directly from the simulation data as a piecewise function on $(\chi,m)$ cells, with probability in the $(b,c)$ cell,
\begin{equation}\label{eq:pcorr}
	p_{{\rm corr},(b,c)} =
	\frac{1}{\Delta\chi_{c}}
	\frac{N_{(b,c)}}{N_{b}},
\end{equation}
where $N_{(b,c)}$ is the number of simulated BHs in the cell, $N_{b}$ the number in the mass bin, and $\Delta\chi_{c}$ the spin range of the cell. 
The two estimators are complementary. \textsc{Beta} imposes a smooth parametric shape and is therefore stable against the finite number of simulated BHs per bin, at the cost of not resolving departures from that shape. \textsc{Bin} follows the simulated distribution faithfully, including such departures, at the cost of a higher variance. Using both estimators improves the reliability of our results.

We considered three comparative models, we call \textsc{PE}, \textsc{Default-bbh}, and \textsc{Merg}. The \textsc{PE} comparative model is the parameter-estimation prior itself, $p_{\rm PE}=\pi_{\rm PE}$, for which the denominator of Eq.~(\ref{eq:bayes}) equals unity and $\mathcal{B}^{\rm PE}_{\chi|m}$ isolates the conditional spin--mass prediction against a population-uninformative baseline.
\textsc{Default-bbh} is the fiducial LVK description of the whole BBH population over GWTC-5.0 \citep{2026arXiv260527226T}, which draws both spins from a common truncated Gaussian and carries no mass dependence, 
\begin{equation}\label{eq:def}
	p_{\rm Def}(\bm{\chi}) = \int {\rm d}\Lambda\,p(\Lambda\mid D)
	\prod_{i=1,2}\mathcal{N}_{[0,1]}(\chi_i;\mu_\chi,\sigma_\chi),
\end{equation}
with $\Lambda=(\mu_\chi,\sigma_\chi)$ marginalized over the LVK hyperposterior inferred from GWTC-5.0 itself; it is therefore a population-informed posterior predictive. 
\textsc{Merg} captures the characteristic spin-magnitude signature
of the hierarchical-merger channel. Both components of GW231123 lie far above the second-generation mass floor, so we assume that both are merger remnants, 
$p_{\rm Merg}(\bm{\chi}) = p_{\rm 2g}(\chi_1)\, p_{\rm 2g}(\chi_2)$,
where $p_{\rm 2g}$ is a Beta function, ${\rm Beta}(6.6, 3.4)$, of mode $0.7$ and concentration $\nu=10$, which reproduces the near-universal remnant spin of a comparable-mass merger
\citep{2008ApJ...684..822B,2017ApJ...840L..24F}, consistent with the spin distribution expected for hierarchical-merger populations \citep{2021NatAs...5..749G,2025ApJ...987..146B}. 
It is deliberately not a complete hierarchical-formation model, so that mass-ratio, redshift and spin-orientation predictions do not enter $p_{\rm Merg}$.  
The $0.7$ mode is conservative given that the GW231123 primary requires more than one merger, assuming the pair-instability mass gap holds. Higher remnant spins require aligned progenitor spins, which are not expected for dynamically assembled BHs.

\begin{figure}[tbp]
	\centering
	\includegraphics[width=\columnwidth]{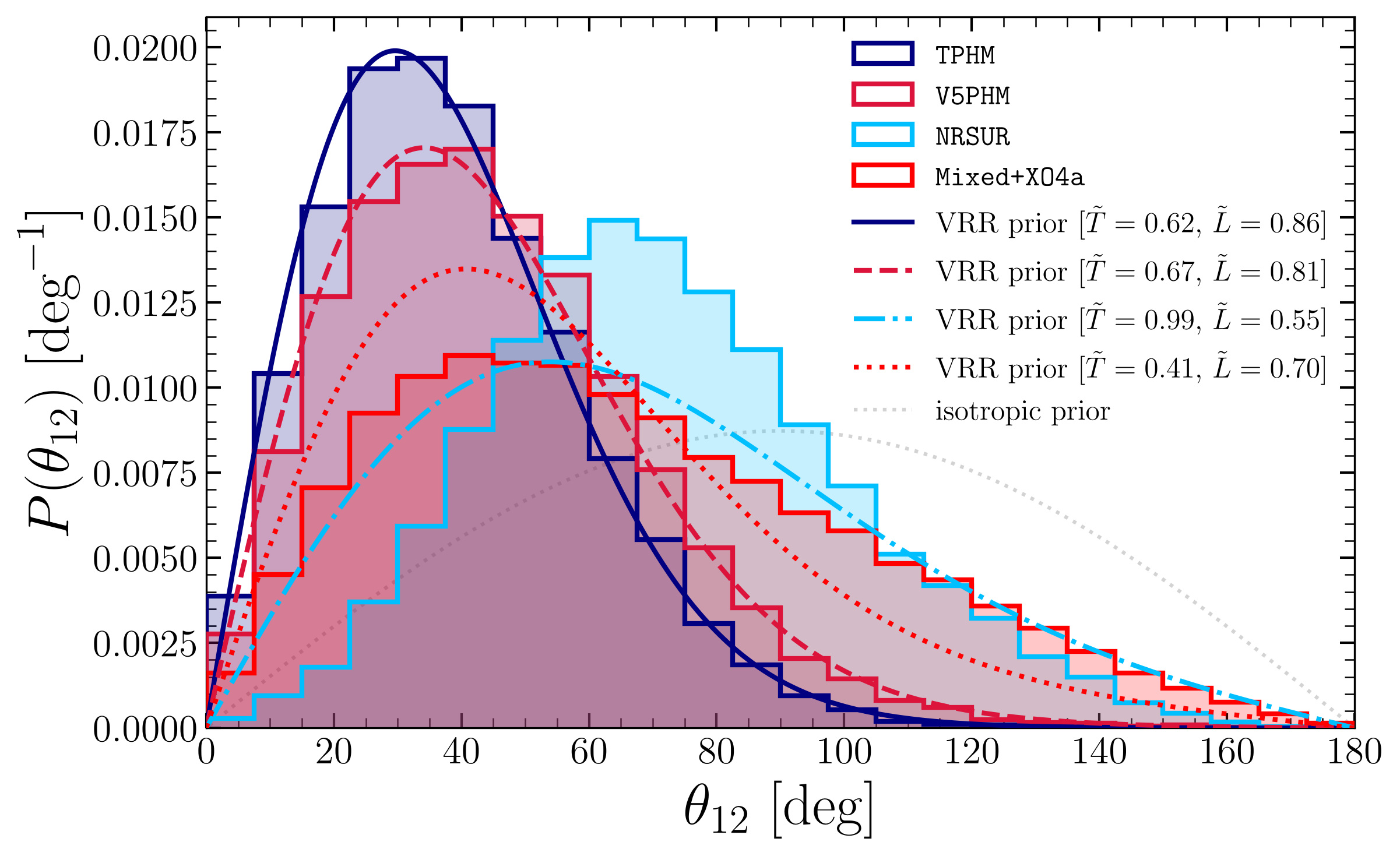}
	\caption{Posterior distribution of the relative spin tilt $\theta_{12}$ for GW231123, for three waveform models and the LVK combined posterior, overlaid with the corresponding best-fit $\theta_{12}$ distributions predicted by VRR, and the LVK isotropic prior.}
	\label{fig:gw231123_joint_theta12_vrr}
\end{figure}

\section{Spin tilt: VRR}\label{sec:vrr}

In brief, VRR \citep{1996NewA....1..149R, 2017ApJ...842...90R} is a form of gravitational relaxation in which the angular momentum vectors of nearly-Keplerian orbits (the angular orientations of orbital planes) reach a statistical ``thermodynamic'' equilibrium independently from the rest degrees of freedom; namely, the magnitudes of the orbital angular momentum (eccentricities) and energy (semi-major axes). \cite{2017ApJ...842...90R} calculated those VRR equilibrium states and identified gravitational phase transitions between ordered and disordered phases. Those are manifested as different geometrical configurations of the orbital spin orientations of the bodies (e.g., stellar BHs) subject to VRR. 
In the ordered phase, the orbit normals (i.e., orbital angular-momentum vectors of the bodies subject to VRR) are concentrated around a common axis, so the orbits form a disk in physical space with thickness which grows with the VRR temperature. 
Above the transition temperature the spread of the normals is so large that the configuration is better described as a thick disk or a nearly isotropic distribution, depending on the total available angular momentum.
VRR operates more effectively in the presence of a central massive object (e.g., a supermassive BH in a nuclear star cluster) that dominates the gravitational potential.
However, it has been further theoretically justified that VRR can operate in the cores of sufficiently compact star clusters even in the absence of a central massive BH \citep{2019ApJ...878..138M, 2020JPhA...53d5002R}. Therefore, in our system, it is possible that VRR operates either if an IMBH ($\sim 10^3\,{\rm M}_{\odot}$) is formed along with the less massive BHs ($\sim 10^2\,{\rm M}_{\odot}$) during the rapid gas-accretion mechanism of \cite{2025A&A...702A.208R} or even if an IMBH failed to form, corresponding to the case of less compact clusters.

In either case, we can assume that the BHs undergoing accretion and growing to masses on the order of $\sim 10^2{\rm M}_{\odot}$ are subject to VRR. 
We further assumed that the individual distances, $r_{\bullet}$, of the BHs from the cluster center are spread over a relatively narrow strip, $\Delta r_{\bullet} \ll \bar{r}_{\bullet}$, around their mean value, $\bar{r}_{\bullet}$, and the individual BH masses, $m_{\bullet}$, do not vary significantly around their mean mass, $\bar{m}_{\bullet}$ (i.e., the so-called one-component system, as described in Section 3 of \citealt{2017ApJ...842...90R}). In our case of interest, this should refer to BH masses on the order of $\bar{m}_{\bullet} \sim 10^2\,{\rm M}_{\odot}$.
The probability density function of the VRR-equilibrium inclinations, $\theta$, of the BH orbital planes (Eq.~(42) in \citealt{2017ApJ...842...90R} for the axially symmetric case) is then
\begin{equation}\label{eq:p_vrr}
	p_{\rm VRR} (\cos\theta) = \frac{1}{Z}e^{\frac{3Q}{2\tilde{T}}q(\cos\theta) + \tilde{\gamma} \cos\theta},\; 
	q(\cos\theta) \equiv \cos^2\theta - \frac{1}{3},
\end{equation}
and $Z \equiv \int_{-1}^{1} \exp\{ \frac{3Q}{2\tilde{T}}q(s) + \tilde{\gamma} s \}\, {\rm d}s$. The VRR temperature, $\tilde{T}$, is a dimensionless parameter related to the Lagrange multiplier, $\beta$, of the entropy variation (Eq.~(19) in \citealt{2017ApJ...842...90R}), translates to the energy, $\tilde{T} = 1/\beta J N_{\bullet}$, where $N_{\bullet}$ is the number of BHs subject to this VRR process.
The quadrupole coupling is $J \sim \frac{3}{8} G \bar{m}_{\bullet}^2/\bar{r}_{\bullet}$. 
The parameter $\tilde{\gamma}$ is related to the Lagrange multiplier of the entropy variation, conjugate to the $z$-axis component of the total angular-momentum of the BH orbits, $L_{z}$. The dimensionless expression $\tilde{L}_z \equiv L_z / N_{\bullet} \ell_{\bullet} \leq 1$, where $\ell_{\bullet} = \bar{m}_{\bullet} \sqrt{G M_{\rm cl}(r < \bar{r}_{\bullet}) \bar{r}_{\bullet}}$, is subject to a self-consistency equation: $\tilde{L}_z = \langle \cos\theta \rangle$. The quadrupole moment satisfies its own self-consistency equation: $Q = \langle q \rangle$, averaged over (\ref{eq:p_vrr}). 
Such self-consistency equations are solved numerically by iterating over the unknown parameter (e.g., $Q$ for the second equation) and calculating at each iteration, the respective mean by numerical integration, $\langle q \rangle = \int_{-1}^{1} p_{\rm VRR}(s)\, q(s)\, {\rm d}s$, then selecting the value of the parameter that coincides with the mean within some numerical tolerance, that is $\langle q \rangle = Q$.
In practice, we calculated the VRR equilibrium $-$ the probability distribution of inclinations of BH orbital planes expressed via Eq.~(\ref{eq:p_vrr}) $-$ by solving the two self-consistency equations for a given pair $\tilde{T}, \tilde{L}_{z}$. This gives the remaining unknowns: $Q$, $\tilde{\gamma}$. Iterating over several values $\tilde{T}, \tilde{L}_{z}$, we obtained a series of several equilibria, which we used to relate to the observed relative BH tilts.

The BH spin will tend strongly to be anti-aligned along the orbital angular momentum direction \citep{2026A&A...709A.5R,2026arXiv_Roupas}. As the BH's sphere of influence is advected along the BH trajectory, the transverse velocity shear injects angular momentum driving the formation of an accretion disk on the orbital plane. Therefore, each BH spin orientation gets determined by the VRR-driven orbital-plane orientation. 

\begin{figure}[tbp]
	\centering
	\includegraphics[width=0.95\columnwidth]{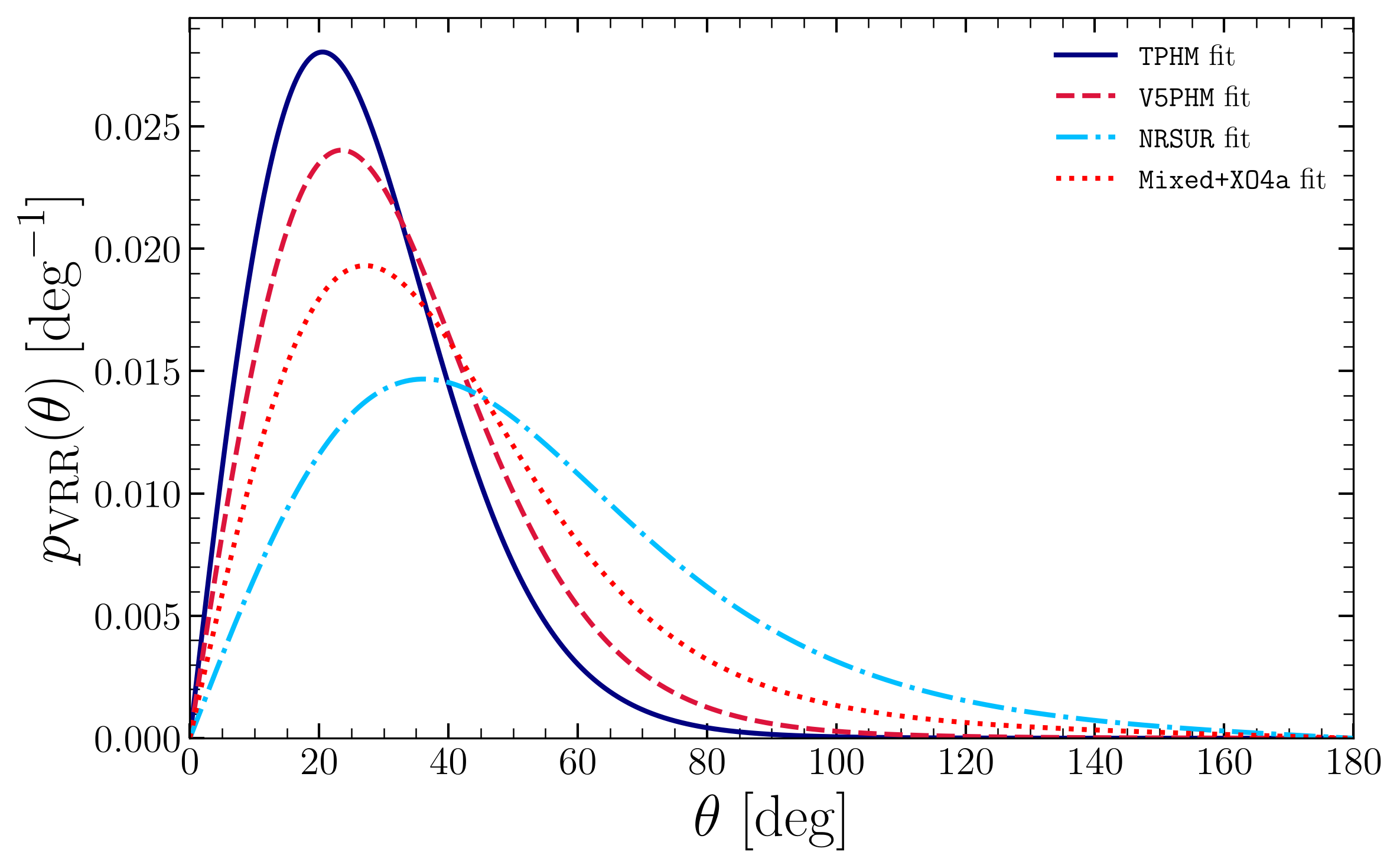}
	\caption{VRR equilibrium distribution, $p_{\rm VRR}$, of the orbital inclination, $\theta$, for the best-fit models of Fig.~\ref{fig:gw231123_joint_theta12_vrr}.}
	\label{fig:theta_pdf_overlay}
\end{figure}

We calculated the VRR prior, which gives a probability density of the relative spin alignment that offers a better fit to the LVK posterior density function $P(\theta_{12})$, as shown in Figure~\ref{fig:gw231123_joint_theta12_vrr}. Remarkably, the models \texttt{TPHM} and \texttt{V5PHM} are fit excellently and even the combined \texttt{Mixed+XO4a} displays a very good fit as well, while \texttt{NRSUR} presents a fairly good fit as it is closer to the isotropic prior. Nevertheless, all four models suggest a similar VRR state with VRR-temperature and total-angular momentum midpoints and half-ranges of $\tilde{T} = 0.70\pm 0.29$, $\tilde{L}_{z} = 0.71\pm 0.15$. 
For all, \texttt{TPHM}, \texttt{V5PHM}, \texttt{NRSUR}, \texttt{Mixed+XO4a}, the values of the pair $\{Q, \tilde{\gamma}\}$ are respectively, $\{0.43, 5.68\}$, $\{0.36, 4.29\}$, $\{0.15, 1.95\}$, $\{0.27, 2.53\}$. The VRR distributions of the orbital orientation, from which the distributions of relative spin tilt of Figure~\ref{fig:gw231123_joint_theta12_vrr} were derived. These are depicted in Figure ~\ref{fig:theta_pdf_overlay}. The VRR spin-orientation Bayes factor is
\begin{equation}
	\mathcal{B}_{\rm VRR}
	=
	\frac{P(d\mid \mathcal{H}_{\rm VRR})}
	{P(d\mid \mathcal{H}_{\rm LVK}^{\theta_{12}})}
	=
	\left\langle
	\frac{P_{\rm VRR}(\theta_{12})}
	{\pi_{\rm LVK}(\theta_{12})}
	\right\rangle_{\rm LVK},
	\label{eq:bayes_vrr}
\end{equation}
where $\pi_{\rm LVK}(\theta_{12})=\tfrac12\sin\theta_{12}$ is the LVK isotropic prior. For the waveform models, \texttt{TPHM}, \texttt{V5PHM}, \texttt{NRSUR}, \texttt{Mixed+XO4a}, we find $\ln \mathcal{B}_{\rm VRR} = 1.29,\, 1.02,\, 0.20,\, 0.46$, respectively. The first two models with an LVK median relative tilt at $\sim 35^o-45^o$ present supporting evidence in favor of VRR over an isotropic prior. The models \texttt{NRSUR}, \texttt{Mixed+XO4a} with an LVK median relative tilt at $\sim 70^o$, $61^o$, which are much closer to the isotropic peak at $90^o$, present only weak evidence.

In Tables~\ref{tab:analysis} and \ref{tab:comparators}, we report the joint Bayes factor of the spin-mass correlation prediction and the VRR hypothesis,
\begin{equation}
	\mathcal{B}
	=
	\left\langle
	\frac{
		p_{\rm corr}(\chi_1\mid m_1)\,
		p_{\rm corr}(\chi_2\mid m_2)}
	{
		\pi_{\rm LVK}(\chi_1)\,
		\pi_{\rm LVK}(\chi_2)}
	\;
	\frac{P_{\rm VRR}(\theta_{12})}
	{\pi_{\rm LVK}(\theta_{12})}
	\right\rangle_{\rm LVK}.
	\label{eq:bayes_joint}
\end{equation}
We find that 
$
	\ln\mathcal{B} \;\approx\; \ln\mathcal{B}_{\chi|m} + \ln\mathcal{B}_{\rm VRR},
	\label{eq:bayes_additivity}
$
within $\lesssim 0.11$, that is the spin-magnitude and the tilt weights of Eq.~(\ref{eq:bayes_joint}) are essentially uncorrelated over the posterior samples. The two hypotheses are therefore constrained independently by the data. Evidence against VRR would not imply evidence against the spin-mass correlation and vice versa. 
We report the joint Bayes factor Eq.~(\ref{eq:bayes_joint}) in the following and we discuss the results of our analysis in detail in Sect.~\ref{sec:results}.

\begin{figure}[tbp]
	\centering
	\includegraphics[width=0.9\columnwidth]{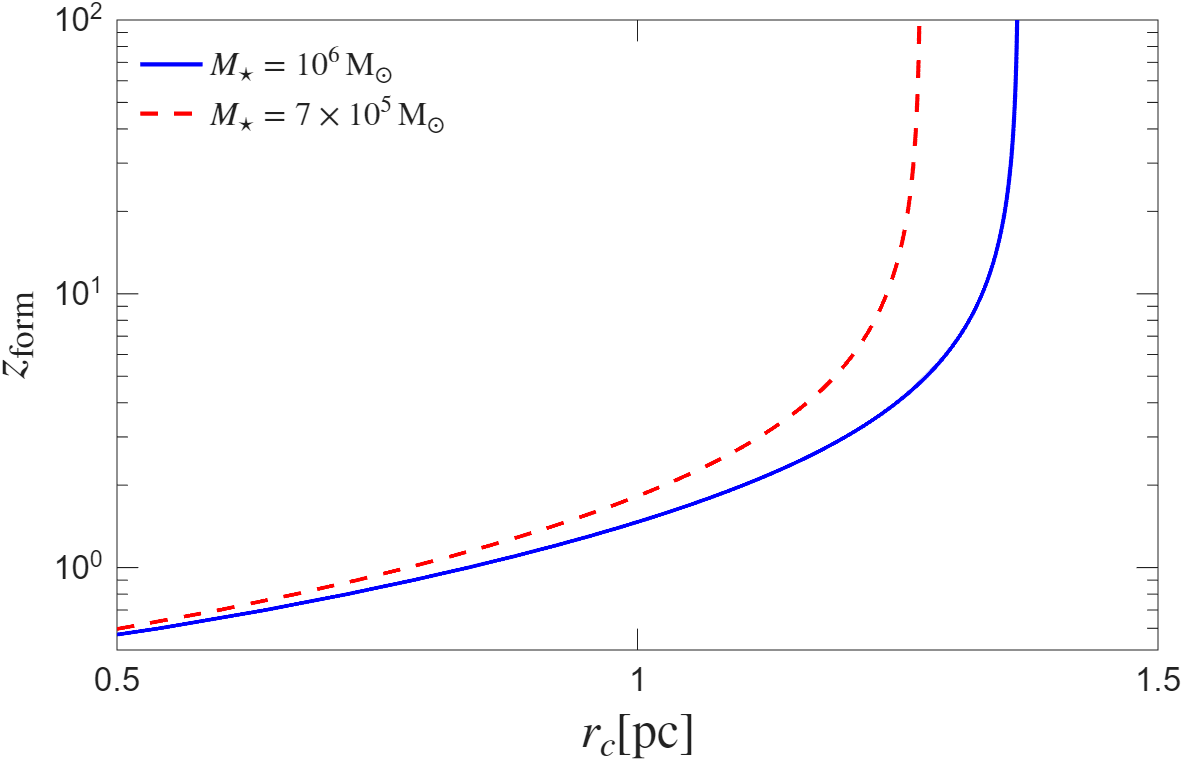}
	\caption{Star-cluster/BBH formation redshift $z_{\rm form}$, versus core radius, $r_c$, for the two cluster masses of Table \ref{tab:analysis}, assuming merger at $z=0.40$.}
	\label{fig:zform_rc}
\end{figure}

\begin{table*}[tb]
	\caption{\label{tab:analysis}
		Logarithm of the Bayes factor against the \textsc{PE} comparative model for several clusters and waveform models.}
	\centering
	\begin{small}
	\begin{tabular}{c | c c c || c c c c c | c c c c c || l }  
		\multicolumn{4}{c||}{Cluster}
		&
		\multicolumn{5}{c|}{$\ln \mathcal{B}^{\rm PE}_{\chi | m}$ [spin-mass correlation]}
		&
		\multicolumn{5}{c||}{$\ln \mathcal{B}^{\rm PE}$ [Joint]}
		&
		\\
		\midrule 
		$M_{\star}$
		&                               
		$r_{c,\star}$
		&                  
		$C_{\rm ini}$
		&     
		$Z$
		&
		\texttt{TPHM}
		&
		\texttt{V5PHM}
		&                       
		\texttt{NRSUR}
		&
		Pool
		&
		\texttt{M+X}
		&
		\texttt{TPHM}
		&
		\texttt{V5PHM}
		&
		\texttt{NRSUR}
		&
		Pool
		&
		\texttt{M+X}
		&
		Est.
		\\[0.2ex]
		\toprule
		\multirow{6}{*}{$10^6$}
		&
		\multirow{2}{*}{$1.0$}
		&
		\multirow{2}{*}{$44.0$}
		&
		\multirow{2}{*}{$0.01$}
		&
		2.02
		&
		1.80
		&
		1.99
		&
		1.94
		&
		1.72
		&
		3.32
		&
		2.83
		&
		2.18
		&
		2.88
		&
		2.26
		&
		\textsc{Beta}
		\\
		&
		&
		&
		&
		2.52
		&
		2.21
		&
		2.53
		&
		2.43
		&
		2.14
		&
		3.81
		&
		3.23
		&
		2.72
		&
		3.35
		&
		2.70
		&
		\textsc{Bin}
		\\[0.5ex]
		&
		\multirow{2}{*}{$1.2$}
		&
		\multirow{2}{*}{$40.0$}
		&
		\multirow{2}{*}{$0.01$}
		&
		2.07
		&
		1.85
		&
		2.01
		&
		1.98
		&
		1.75
		&
		3.36
		&
		2.87
		&
		2.21
		&
		2.92
		&
		2.30
		&
		\textsc{Beta}
		\\
		&
		&
		&
		&
		2.43
		&
		2.13
		&
		2.38
		&
		2.33
		&
		2.05
		&
		3.72
		&
		3.16
		&
		2.58
		&
		3.26
		&
		2.61
		&
		\textsc{Bin}
		\\[0.5ex]
		&
		\multirow{2}{*}{$1.0$}
		&
		\multirow{2}{*}{$44.0$}
		&
		\multirow{2}{*}{$0.10$}
		&
		1.94
		&
		1.71
		&
		1.82
		&
		1.83
		&
		1.63
		&
		3.23
		&
		2.73
		&
		2.02
		&
		2.77
		&
		2.16
		&
		\textsc{Beta}
		\\
		&
		&
		&
		&
		2.40
		&
		2.10
		&
		2.31
		&
		2.28
		&
		2.02
		&
		3.69
		&
		3.12
		&
		2.50
		&
		3.22
		&
		2.57
		&
		\textsc{Bin}
		\\[0.5ex]
		\midrule
		\multirow{8}{*}{$7\times 10^5$}
		&
		\multirow{2}{*}{$0.7$}
		&
		\multirow{2}{*}{$44.0$}
		&
		\multirow{2}{*}{$0.01$}
		&
		2.10
		&
		1.89
		&
		2.05
		&
		2.02
		&
		1.80
		&
		3.40
		&
		2.92
		&
		2.26
		&
		2.96
		&
		2.35
		&
		\textsc{Beta}
		\\
		&
		&
		&
		&
		2.02
		&
		1.81
		&
		2.04
		&
		1.96
		&
		1.75
		&
		3.32
		&
		2.85
		&
		2.25
		&
		2.90
		&
		2.29
		&
		\textsc{Bin}
		\\[0.5ex]
		&
		\multirow{2}{*}{$0.9$}
		&
		\multirow{2}{*}{$35.0$}
		&
		\multirow{2}{*}{$0.01$}
		&
		1.92
		&
		1.72
		&
		1.97
		&
		1.88
		&
		1.65
		&
		3.22
		&
		2.76
		&
		2.17
		&
		2.80
		&
		2.19
		&
		\textsc{Beta}
		\\
		&
		&
		&
		&
		2.20
		&
		1.94
		&
		2.32
		&
		2.17
		&
		1.91
		&
		3.50
		&
		2.99
		&
		2.53
		&
		3.08
		&
		2.46
		&
		\textsc{Bin}
		\\[0.5ex]
		&
		\multirow{2}{*}{$0.7$}
		&
		\multirow{2}{*}{$44.0$}
		&
		\multirow{2}{*}{$0.10$}
		&
		2.60
		&
		2.31
		&
		2.59
		&
		2.51
		&
		2.21
		&
		3.90
		&
		3.34
		&
		2.79
		&
		3.44
		&
		2.79
		&
		\textsc{Beta}
		\\
		&
		&
		&
		&
		2.75
		&
		2.41
		&
		2.79
		&
		2.66
		&
		2.34
		&
		4.03
		&
		3.43
		&
		2.98
		&
		3.58
		&
		2.91
		&
		\textsc{Bin}
		\\[0.5ex]
		&
		\multirow{2}{*}{$0.8$}
		&
		\multirow{2}{*}{$37.5$}
		&
		\multirow{2}{*}{$0.10$}
		&
		2.09
		&
		1.88
		&
		2.09
		&
		2.02
		&
		1.77
		&
		3.38
		&
		2.91
		&
		2.29
		&
		2.96
		&
		2.33
		&
		\textsc{Beta}
		\\
		&
		&
		&
		&
		2.16
		&
		1.91
		&
		2.25
		&
		2.12
		&
		1.84
		&
		3.45
		&
		2.96
		&
		2.44
		&
		3.03
		&
		2.40
		&
		\textsc{Bin}
		\\[0.5ex]
		\bottomrule
	\end{tabular}
\end{small}
	\tablefoot{$M_{\star}$: cluster stellar mass in ${\rm M}_{\odot}$;
		$r_{c,\star}$: cluster core radius in pc; $C_{\rm ini}$: initial cluster
		compactness before gas depletion; $Z$: metallicity in ${\rm Z}_{\odot}$;
		Pool: equal-weight pooling of posterior samples from \texttt{TPHM},
		\texttt{V5PHM} and \texttt{NRSUR}; \texttt{M+X}: the \texttt{Mixed+XO4a}
		combination of all five waveform models released by the LVK. The last column
		gives the estimator of $p_{\rm corr}(\chi\mid m)$ of each row,
		Eq.~(\ref{eq:pcorr}) for \textsc{Bin}. $\ln\mathcal{B}^{\rm PE}_{\chi|m}$:
		spin-mass correlation Bayes factor, Eq.~(\ref{eq:bayes});
		$\ln\mathcal{B}^{\rm PE}$: joint spin-mass and VRR Bayes factor,
		Eq.~(\ref{eq:bayes_joint}).}
\end{table*}

\section{Formation time}\label{sec:time}

First, we must consider whether there is enough time since the BBH formation for the BHs to merge at $z=0.40$, as estimated by LVK for the GW231123 \citep{2025ApJ...993L..25A}. On the other hand, we consider whether it is possible that our proposed mechanism is so effective that the merger occurs so rapidly at high $z$, that is, long before $z=0.40$. In this context, we have to estimate the age of the proto-cluster so that it can produce a GW231123 type of merger at $z=0.40$.

We assume that the BBH is formed rapidly, within less than $\sim 10\,{\rm Myr}$ following gas depletion, that is within $\sim 20\,{\rm Myr}$ since the first star formation event of the proto-cluster. This is justified by recent simulations \citep{2026ApJ...998..138M,2025MNRAS.538..639B}. Theoretically, assuming BBH formation via 3-body encounters, we get a formation timescale of $t_{3\rm b,BBH} \approx 1 \, {\rm Myr}\,(n/10^5{\rm pc}^{-3})^{-2}(\sigma/10{\rm km}/{\rm s})^{9} (m_1 / 100{\rm M}_{\odot})^{-5} $ \citep{1995MNRAS.272..605L,2020PhRvD.102d3002B}. Therefore, we can identify the proto-cluster age, the BHs formation lookback time and the BBH formation lookback time. All these events occur within a window $\lesssim 50\,{\rm Myr}$, small compared to the lookback time, $\sim 4{\rm Gyr}$, at $z=0.40$, as well as the times at even higher $z$.
 
 We calculated the BBH-merger timescale, $\tau$, according to the procedure described in Appendix~\ref{app:time}.
 In Figure~\ref{fig:zform_rc}, we depict the redshift of the proto-cluster birth, $z_{\rm form}$, identified as well with the BBH-formation approximately, with respect to the cluster size. In calculating $z_{\rm form}$, we assumed that the merger occurred at $z=0.40$. We added the merger timescale, $\tau$, to the corresponding lookback time and then converted the new lookback time to redshift. 
 
 We found realistic possible formation times of the proto-cluster. The older possible formation time is during the re-ionization era, $z_{\rm form} \sim 6-10$, also likely pointing to the low-metallicity case. Therefore, proto-clusters such as those observed by the James Webb Space Telescope (JWST) in the Cosmic Gems arc galaxy at $z=10.2$ \citep{2024Natur.632..513A}, are plausible realistic candidates. Nevertheless, the more plausible formation time, inferred by clusters of sizes $r_c = 0.7-1.2\,{\rm pc}$, corresponds to $z_{\rm form} \sim 1-3$, possibly favoring the sub-solar metallicity case.
 
\begin{table}[tb]
	\caption{\label{tab:comparators}
		Logarithm of the spin-mass and joint Bayes factors of the accretion channel
		against the three comparative models, for the favored star cluster.}
	\centering
	\begin{small}
	\begin{tabular}{l || c c c | c c c || l }
		&
		\multicolumn{3}{c|}{$\ln \mathcal{B}_{\chi | m}$}
		&
		\multicolumn{3}{c||}{$\ln \mathcal{B}$ [Joint]}
		&
		\\
		\midrule
		Model
		&
		\textsc{PE}
		&
		\textsc{Def}
		&
		\textsc{Merg}
		&
		\textsc{PE}
		&
		\textsc{Def}
		&
		\textsc{Merg}
		&
		{\footnotesize Est}
		\\[0.2ex]
		\toprule
		\multirow{2}{*}{\texttt{TPHM}}
		&
		2.60
		&
		5.65
		&
		2.69
		&
		3.90
		&
		6.95
		&
		3.99
		&
		{\footnotesize \textsc{Beta}}
		\\
		
		&
		2.75
		&
		5.79
		&
		2.84
		&
		4.03
		&
		7.08
		&
		4.12
		&
		{\footnotesize \textsc{Bin}}
		\\[0.5ex]
		\multirow{2}{*}{\texttt{V5PHM}}
		&
		2.31
		&
		4.99
		&
		2.21
		&
		3.34
		&
		6.02
		&
		3.23
		&
		{\footnotesize \textsc{Beta}}
		\\
		
		&
		2.41
		&
		5.09
		&
		2.31
		&
		3.43
		&
		6.11
		&
		3.33
		&
		{\footnotesize \textsc{Bin}}
		\\[0.5ex]
		\multirow{2}{*}{\texttt{NRSUR}}
		&
		2.59
		&
		5.54
		&
		2.61
		&
		2.79
		&
		5.74
		&
		2.81
		&
		{\footnotesize \textsc{Beta}}
		\\
		
		&
		2.79
		&
		5.74
		&
		2.81
		&
		2.98
		&
		5.94
		&
		3.00
		&
		{\footnotesize \textsc{Bin}}
		\\[0.5ex]
		\multirow{2}{*}{Pool}
		&
		2.51
		&
		5.39
		&
		2.51
		&
		3.44
		&
		6.32
		&
		3.44
		&
		{\footnotesize \textsc{Beta}}
		\\
		
		&
		2.66
		&
		5.54
		&
		2.66
		&
		3.58
		&
		6.46
		&
		3.57
		&
		{\footnotesize \textsc{Bin}}
		\\[0.5ex]
		\multirow{2}{*}{\texttt{M+X}}
		&
		2.21
		&
		4.44
		&
		2.06
		&
		2.79
		&
		5.01
		&
		2.63
		&
		{\footnotesize \textsc{Beta}}
		\\
		
		&
		2.34
		&
		4.56
		&
		2.18
		&
		2.91
		&
		5.13
		&
		2.76
		&
		{\footnotesize \textsc{Bin}}
		\\[0.5ex]
		\bottomrule
	\end{tabular}
	\end{small}
	\tablefoot{Favored cluster: $M_{\star} = 7\times10^{5}\,{\rm M}_{\odot}$,
		$r_{c,\star} = 0.7$ pc, $C_{\rm ini} = 44.0$, $Z = 0.1\,{\rm Z}_{\odot}$.
		\textsc{PE}, \textsc{Def}, and \textsc{Merg} denote the comparative models
		of Sect.~\ref{sec:spin-mass}: the parameter-estimation prior, the
		\textsc{Default-bbh} population predictive of Eq.~(\ref{eq:def}), and the
		merger-remnant model. The last column gives the estimator of
		$p_{\rm corr}(\chi\mid m)$ of each row.}
\end{table}

\section{Results}\label{sec:results}

Table~\ref{tab:analysis} reports the spin--mass and joint Bayes factors against the \textsc{PE} comparative model for all seven cluster models, the five waveform sets, and both estimators of $p_{\rm corr}(\chi\mid m)$. The spin--mass factor spans $\ln\mathcal{B}^{\rm PE}_{\chi|m} = 1.6-2.8$, positive evidence throughout \citep{Kass01061995}. 
The joint factor spans $\ln\mathcal{B}^{\rm PE} = 2.0-4.0$; for \texttt{TPHM} it exceeds $3.2$ in every cluster, which offers strong evidence, and for the Pool, it reaches $3.4-3.6$ in the favored cluster. 
The Monte Carlo uncertainty is $\leq 0.02$ throughout, and the joint factor agrees with the sum $\ln\mathcal{B}^{\rm PE}_{\chi|m} + \ln\mathcal{B}_{\rm VRR}$ within $\leq 0.11$, so the two hypotheses are effectively independent.

The spread across waveform sets is $\leq 0.4$ for the retained models and follows the informativeness of the secondary spin. 
The medians all lie inside the predicted plateau, $\chi_2 = 0.80-0.91$, while the credible widths differ substantially, $\Delta_{90}\chi_2 = 0.39-0.83$: the broadest posteriors, \texttt{V5PHM} and \texttt{M+X}, give the lowest factors, while
\texttt{TPHM} and \texttt{NRSUR}, with similar widths, agree within $0.12$. 
The width of the secondary spin therefore drives the spread and the median position plays a secondary role. The two models absent from the table continue this trend: \texttt{XPHM} and \texttt{XO4a} give $\ln\mathcal{B}^{\rm PE}_{\chi|m} = 0.9-1.3$ and $1.1-1.5$, respectively, positive evidence weakened by their uninformative (prior-dominated) secondary spins.

The spread across clusters, $\ln\mathcal{B}^{\rm PE}_{\chi|m} = 1.6-2.6$ for \textsc{Beta}, tracks the form of the predicted spin--mass relation.
Across the clusters of Table~\ref{tab:analysis} the median simulated spin saturates at maximum values of $0.84-0.89$, and the rise toward the plateau occurs around transition masses of $40-56\,{\rm M}_{\odot}$. 
The favored cluster, $M_{\star} = 7\times10^{5}\,{\rm M}_{\odot}$, $r_{c} = 0.7$ pc, $Z = 0.1\,{\rm Z}_{\odot}$, combines the lowest transition mass, $\simeq 40\,{\rm M}_{\odot}$, with the sharpest rise and a high maximum spin, $0.88$, so that both GW231123 components lie deep inside the high-spin plateau where the predicted density concentrates. The \textsc{Bin} and \textsc{Beta} estimators agree on the log Bayes factors within $0.6$ providing confidence in the reported values as independent methods of estimating the conditional probability distribution of the simulation data. The \textsc{Bin} estimator is stable under reasonable changes of the spin-lattice resolution and the minimum cell occupancy.
Since it is also referring directly to the discrete simulation data points, it could be regarded as the more faithful one. It is larger than \textsc{Beta} (which is a continuous fit of the discrete data) for more massive clusters.

Table~\ref{tab:comparators} reports the factors against all three comparative models for the favored cluster. The factors against \textsc{Def} and \textsc{Merg} follow on
$\ln\mathcal{B}^{X} = \ln\mathcal{B}^{\rm PE}
- \ln\langle p_{X}/\pi_{\rm PE}\rangle$, where the last term depends only on the waveform set: adding $3.05$, $2.68$, $2.95$, $2.88$, $2.23$ to the \texttt{TPHM}, \texttt{V5PHM}, \texttt{NRSUR}, Pool, \texttt{M+X} entries of Table~\ref{tab:analysis} gives $\ln\mathcal{B}^{\rm Def}$, and adding $0.09$, $-0.11$, $0.02$, $0.00$, $-0.15$ gives
$\ln\mathcal{B}^{\rm Merg}$.  
The strongest evidence in favor of accretion is obtained against the population-informed model, \textsc{Default-bbh}, reaching $\ln\mathcal{B}^{\rm Def} = 5.0-7.1$ for the joint hypothesis in the favored cluster. 
This states with confidence that the spin magnitudes of GW231123 are not drawn from the processes shaping the bulk of the GWTC-5.0 population; instead, they appear to be consistently accretion-driven.
The factor against \textsc{Merg} yields positive evidence for the accretion channel from the spin magnitudes alone,
$\ln\mathcal{B}^{\rm Merg}_{\chi|m} = 2.0-2.8$, and positive to strong evidence for the joint hypothesis, $\ln\mathcal{B}^{\rm Merg} = 2.6-4.1$, for the favored cluster reported in Table~\ref{tab:comparators}.
Moreover, the joint factor amounts to a direct comparison of the two channels in the spin sector. Isotropic spin orientations are the natural prediction of dynamically assembled mergers entering implicitly Eq.~(\ref{eq:bayes_joint}). Therefore, the tilt information by VRR additionally strengthens the accretion channel against \textsc{Merg} to the spin-mass relation by the full $\ln\mathcal{B}_{\rm VRR}$.

The evidence for the accretion channel is positive to very strong for every cluster, waveform set, estimator, and comparative model considered. 
For \texttt{TPHM}, \texttt{V5PHM}, and the Pool, the joint hypothesis is very strong against the population-informed \textsc{Default-bbh} model in every cluster, while \texttt{TPHM} is strong against the merger comparator \textsc{Merg} in every cluster.
For the favored cluster, the evidence is strong for the joint hypothesis of \texttt{TPHM}, \texttt{V5PHM}, and the Pool, against both the uninformative prior and the merger comparator \textsc{Merg}, as well as against the population-informed \textsc{Default-bbh} comparative model for all waveforms.

\section{Conclusion}\label{sec:conclusion}

We propose a coherent channel for the origin of gravitational-wave events with BH masses within (or above) the theorized mass gap, high spins, and relative spin tilts, similar to GW231123. The formation process of a compact massive star cluster leaves behind a distinct signature on the BHs  generated by stellar evolution and subject to the accretion of residual gas before it gets depleted: a BH spin-mass correlation, and BH spin anti-alignment with BH orbital angular momentum \citep{2026A&A...709A.5R,2026arXiv_Roupas}. A BH subcluster forms that could be subject to VRR (especially if an IMBH is generated), which drives the orbital plane orientations \citep{2017ApJ...842...90R} and, thus, the BH spin alignments.

GW231123 is consistent with these predictions. The spin-mass correlation prediction is by itself independent from the VRR hypothesis. For the pool of the informative waveform models, it alone gives positive Bayesian evidence over the LVK prior and over the characteristic spin signature of the hierarchical-merger channel, \textsc{Merg}, and very strong evidence over the GWTC-5.0 population-informed fiducial model. VRR equilibria reproduce well the observed relative spin tilt (Fig.~\ref{fig:gw231123_joint_theta12_vrr}). 
For the favored cluster model ($M_{\star} = 7\cdot 10^5\,{\rm M}_{\odot}$, $r_c=0.7\,{\rm pc}$, $Z=0.1\,{\rm Z}_{\odot}$), the joint hypothesis gives strong Bayesian evidence over both the LVK prior and \textsc{Merg}, along with very strong evidence over the population-informed model (Table~\ref{tab:comparators}). 
The relative tilt enters the comparison against \textsc{Merg} implicitly, because \textsc{Merg} assumes isotropic spin orientations, while the accretion-driven channel is assigned the VRR equilibrium distribution.

Moreover, we find that the observed merger at $z=0.40$ is consistent with a star cluster formed at a realistic redshift. The more plausible cluster formation redshift is $z_{\rm form} \approx 1-3$, favoring also a mildly sub-solar metallicity. Nevertheless, the oldest allowed clusters trace back to the reionization era $z_{\rm form} \approx 6-10$, allowing for the lower metallicity possibility as well, compatible with the oldest proto-clusters observed to date by the JWST at $z=10.2$ in the Cosmic Gems arc galaxy \citep{2024Natur.632..513A}.

A major future investigation could focus on a population-level test of the accretion-driven channel, especially once a larger sample of high-mass events becomes available. Our comparison against the hierarchical-merger channel is restricted to the spin sector. A population-level test could bring in the mass ratio and the redshift and could allow this channel to be distinguished from other scenarios.

\begin{acknowledgements}
	I thank Monica Colpi and Massimo Dotti for useful discussions.
	ZR is supported by the European Union's Horizon Europe Research and Innovation Programme under the Marie Sk\l{}odowska-Curie grant agreement No.~101149270--ProtoBH. 
\end{acknowledgements}
		 
\bibliography{aa62023-26}
\bibliographystyle{aa_url}	

\begin{appendix}

\section{BBH-merger timescale}\label{app:time}

\begin{figure}[!hb]
	\centering
	\includegraphics[width=0.9\columnwidth]{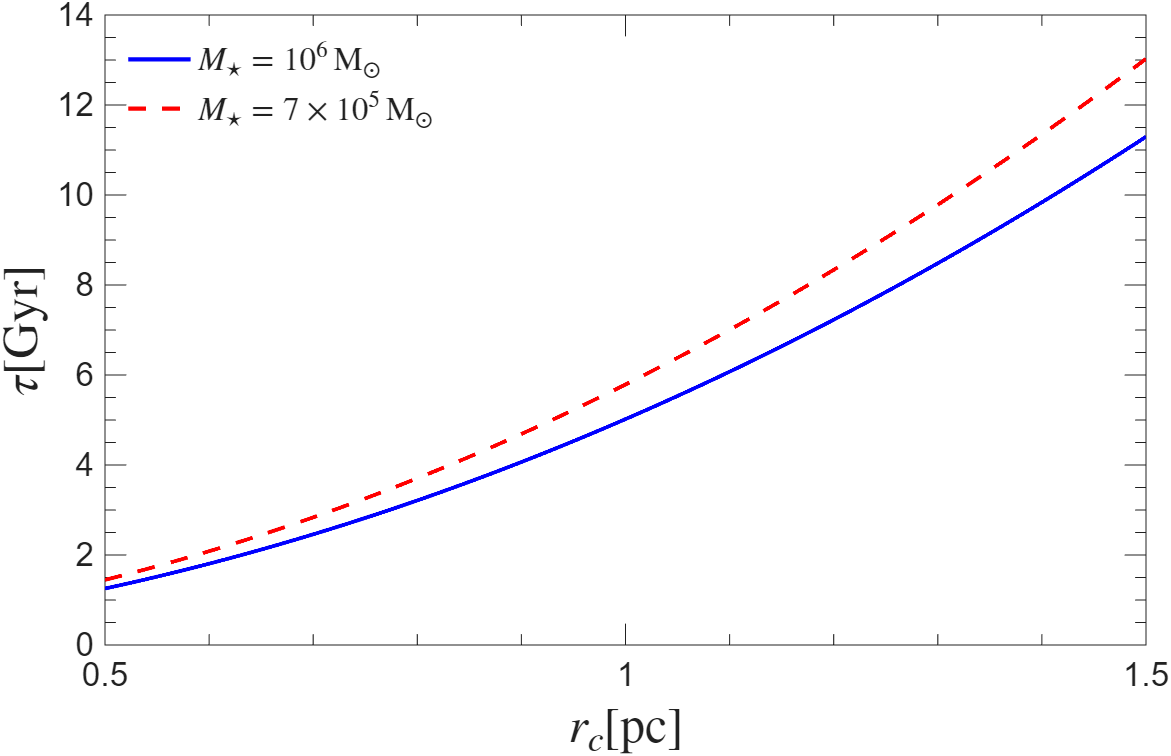}
	\caption{The BBH-merger timescale $\tau$, versus core radius $r_c$, for the two cluster masses of Table \ref{tab:analysis}.}
	\label{fig:tau_rc}
\end{figure}

We estimated the timescale of the GW231123 BBH-merger in the star-cluster environment considered here, expressed as follows. Once the BBH forms it will tend to get harder via three-body encounters if it was formed hard in the first place \citep{1975MNRAS.173..729H, 1975AJ.....80..809H}, that is we assume the initial BBH separation is equal to the threshold for a hard binary, 
\begin{equation}
	a_0 = a_{\rm hard} \equiv G\frac{m_1 m_2}{\left\langle m \right\rangle_\star \sigma_\star^2},
\end{equation}
where $\langle m \rangle_\star$ is the mean stellar mass, $\sigma_\star$ the velocity dispersion of the cluster core, and $m_1$, $m_2$ the masses of the primary and secondary BBH components respectively.
At this scale of separation, the BBH gets hardened primarily via three-body encounters \citep{1996NewA....1...35Q} expressed as
\begin{equation}\label{eq:dota_3b}
	\dot{a}_{\{3{\rm b}\}} = - A\frac{G\rho_\star}{\sigma_\star} a^2,
\end{equation}
where $A = \mathcal{O}(10)$ is a dimensionless constant estimated by numerical experiments,  $a$ is the binary semi-major axis, and $\rho_\star$ the mass density of the core. 
We assume that $\rho_\star$, $\sigma_\star$ are constant.
Beyond some separation threshold, $a_1 = a(t_1)$, the GW-emission hardening starts to dominate over 3-body hardening. 
In the adiabatic approximation, the hardening via GW emission is in the circular-orbit limit \citep{1964PhRv..136.1224P} via
\begin{equation}\label{eq:dota_GW}
	\dot{a}_{\{\rm GW\}} = -\frac{64}{5}\frac{G^3}{c^5} m_1 m_2 (m_1 + m_2) a^{-3}. 
\end{equation}
Eccentricity, which three-body encounters tend to excite, shortens the merger time therefore our estimated BBH-merger timescale is an upper limit in this respect.
The separation $a_1$ can be estimated as
\begin{equation}
\dot{a}_{\{\rm GW\}}(t_1) \sim \dot{a}_{\{3{\rm b}\}}(t_1)
\Rightarrow
a_1 = \left( \frac{64}{5 A}\frac{G^2}{c^5} \frac{\sigma_\star}{\rho_\star} m_1 m_2 (m_1 + m_2)\right)^{1/5}.
\end{equation}
Assuming that 3-body hardening dominates from $t=0$ until $t_1$ and GW emission from $t_1$ until the merger, the BBH-merger timescale is
\begin{equation}
	\tau \approx \frac{\sigma_\star}{A G \rho_\star} \left( \frac{1}{a_1} - \frac{1}{a_0}\right) + 
	\frac{5}{256}\frac{c^5}{G^3}\frac{a_1^4}{m_1 m_2 (m_1 + m_2)},
\end{equation}
depicted in Fig.~\ref{fig:tau_rc}.

\end{appendix}
	 		 	
\end{document}